\documentclass[journal=jacsat,manuscript=article]{achemso}

\usepackage{amsmath,amssymb}
\usepackage{chemformula} %
\usepackage[T1]{fontenc} %

\author{Han Xie}
\affiliation{School of Physics and Key Laboratory of Functional Polymer Materials of Ministry of Education, Nankai University, and Collaborative Innovation Center of Chemical Science and Engineering, Tianjin 300071, China}
\alsoaffiliation{Department of Physics and Astronomy, University of Waterloo, Waterloo, Ontario N2L 3G1, Canada}
\author{Wenyu Liu}
\affiliation{School of Physics and Key Laboratory of Functional Polymer Materials of Ministry of Education, Nankai University, and Collaborative Innovation Center of Chemical Science and Engineering, Tianjin 300071, China}
\author{Zhenyue Lu}
\affiliation{School of Physics and Key Laboratory of Functional Polymer Materials of Ministry of Education, Nankai University, and Collaborative Innovation Center of Chemical Science and Engineering, Tianjin 300071, China}
\author{Jeff Z.Y. Chen}
\email{jeffchen@uwaterloo.ca}
\affiliation{Department of Physics and Astronomy, University of Waterloo, Waterloo, Ontario N2L 3G1, Canada}
\author{Yao Li}
\email{liyao@nankai.edu.cn}
\affiliation{School of Physics and Key Laboratory of Functional Polymer Materials of Ministry of Education, Nankai University, and Collaborative Innovation Center of Chemical Science and Engineering, Tianjin 300071, China}

\title[An \textsf{achemso} demo]
  {Competing Hexagonal and Square Lattices on a Spherical Surface}

\abbreviations{Hex, Sq}
\keywords{Computer simulations; Topological defects; Self assembly; Curvature-induced frustration; counter-domain defects; Gauss-Bonnet theorem}

\begin{document}

\begin{tocentry}

    \includegraphics{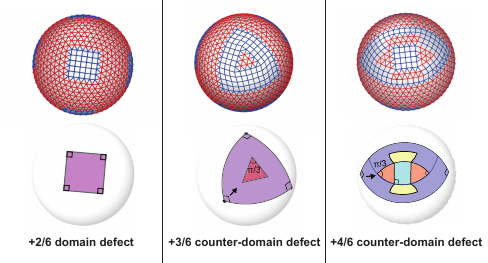}

\end{tocentry}

\begin{abstract}
  The structural properties of packed soft-core particles provide a platform to understand the cross-pollinated physical concepts in solid-state- and soft-matter physics. Confined on spherical surface, the traditional differential geometry also dictates the overall defect properties in otherwise regular crystal lattices. Using molecular dynamics simulation of the Hertzian model as a tool, we report here the emergence of new types of disclination patterns: domain and counter-domain defects, when hexagonal and square patterns coexist. A new angle is presented to understand the incompatibility between tiling lattice shapes and the available spherical areal shapes, which is common in nature---from molecular systems in biology to backbone construction in architectures. 
  
  \textbf{Keywords: Computer simulations; Topological defects; Self assembly; Curvature-induced frustration; counter-domain defects; Gauss-Bonnet theorem}
  
\end{abstract}

\begin{figure*}
	\centering{}\includegraphics[width=1\textwidth]{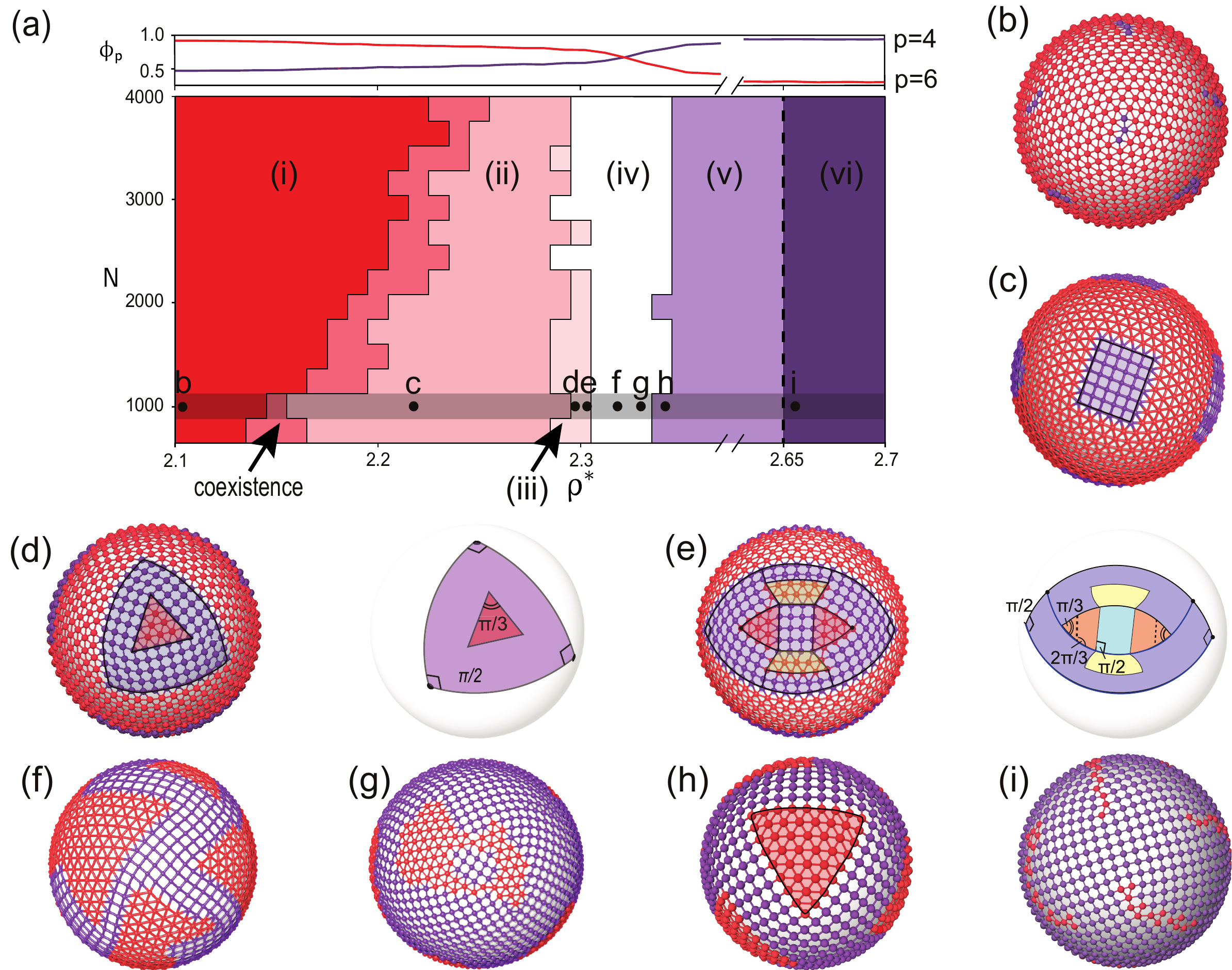}
	\renewcommand{\thefigure}{1}
	\caption{\textbf{State diagram of the Hex-Sq coexistent region and simulation snapshots.} 
Six main regimes, (i) to (vi), are observed from our computer simulations of the Hertzian potential, depicted in (a).
The upper panel of (a) displays the overall lattice order parameter $\phi_4$ and $\phi_6$ for $N=1000$.
The lettered locations in the state diagram (b-i) are where the snapshots in (b) to (i) are taken. In regime-(i),
 deformed Hex lattices on spherical surface contain either localized point disclinations or scars [plot (b)].
In regime-(ii), Sq-lattice domains
appear in the background of
deformed Hex lattices [plots (c)]. In regime-(iii), enclave Hex-lattice domains
exist inside Sq-lattice domains, which are embedded in the
background of
deformed Hex lattices [plots (d) and (e)]. Regime-(iv) contains significant Hex- and Sq-lattice domains interwind together, where no clear
global symmetry can be detected [plots (f) and (g)]. In regime-(v), localized Hex-lattice domains exist in the a majority of Sq-lattice domains [plot (h)].
Beyond that, in (vi), the spherical surface is covered by deformed Sq lattices containing isolated point or scar disclinations [plot (i)]. 
{In between (i) and (ii)}, coexistence of the structures seen in (i) and (ii) is observed.}
	\label{FIG1}
\end{figure*}

\begin{figure}
	\includegraphics[width=0.5\columnwidth]{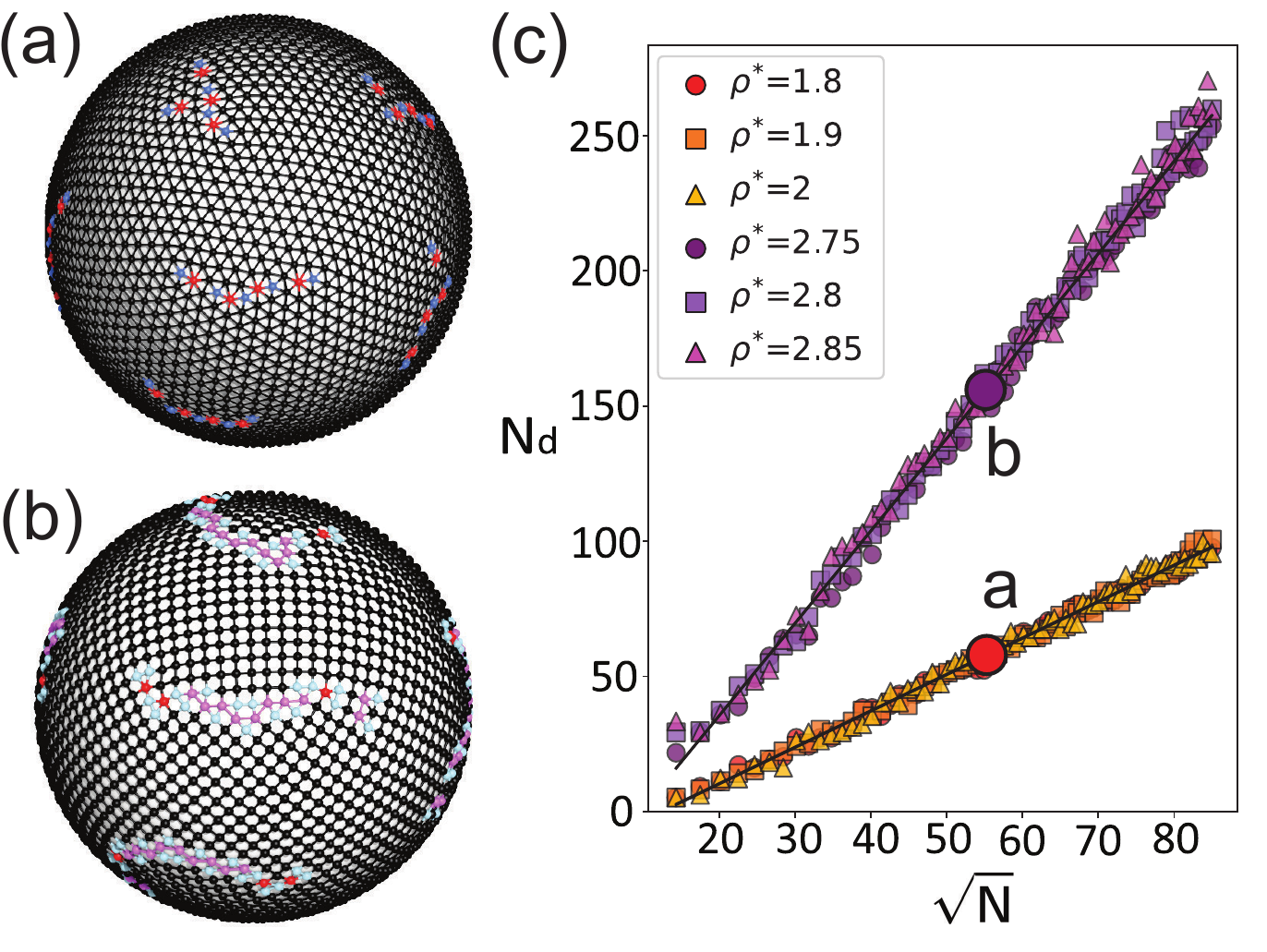}
	\renewcommand{\thefigure}{2}
	\caption{
		\textbf{Properties of scars.}
		Configurational snapshots for scars in both
		Hex lattice  and Sq lattice are shown in (a) and (b),
		where
		$w=-1/6$, $+1/6$, $-1/12$, and $+1/12$ defects are represented by molecules having red, blue, pink, and cyan colors.
		The number of excess disclinations, $N_d$, in both Sq and Hex lattices, is shown in plot ({c}) as a function of $\sqrt{N}$, which clearly demonstrates the
		power law in \eqref{Nddef}, for various values of densities. Letter labels in (c) are the parameter locations of snapshots in (a), (b).
	}
	\label{FIG2}
\end{figure}

Understanding the physical origin of defect features of ordered or partially ordered states at the molecular level
is essential in materials science.
Deformations, dislocations and disclinations can play a significant role in
making matters functional, instead of simply being structural flaws  \cite{Chaikin1995}.
Of recent interest is the assembly of molecules into ordered lattices on curved surfaces of non-trivial
Euler characteristics, which must accompany defects of a fixed total winding number --- a measure of the disclination nature \cite{Nelson2002}. Here, we demonstrate that the frustration between lattice tiling and the spherical
geometry as well as the requirement of potential-energy minimum of the total system, {yields
	domain defects and counter-domain defects [see Fig. \ref{FIG1}], beyond point defects [Fig. S1 in SI] and scars [Fig. \ref{FIG2}]} usually studied before.
This {computer-simulation} study
{presents} a global view of the defect emergence.

In contrast to conventional planar systems where
crystals or liquid crystals display
regularity in molecular-bond orientations,
the spherical topology disrupts the orientational order and forces them to adopt defective
configurations. %
Extensive experimental and theoretical studies
discovered various defect patterns and configurational symmetries formed by molecules of different shapes. Examples include confined molecules or polymer beads ($\sim  10\mu$m) on spherical droplets formed by water-oil emulsions that display local hexagonal crystalline order \cite{Bausch2003,Guerra2018,Das2022,Meng2014}, 
the assembly of divalent nanoparticles \cite{DeVries2007} that exhibits polar order  \cite{Jiang2007},
the structure of double emulsions of liquid-crystal droplets   ($\sim  10\mu$m) \cite{Lopez-Leon2011} that contain local
nematic order  \cite{Bates2008,Shin2008}.
It is also believed that surface
structural morphology are linked to the functionality of
biological matter, such as virus shells which display both
hexagonal  \cite{Roos2010} and square
lattices made of proteins  ($\sim  10$nm) \cite{Laughlin2022,Plevka2008},
bacterial surfaces \cite{Sleytr2014} and vesicles made of
lipids ($\sim  10$nm) \cite{ Mackintosh1991,Hirst2013}, as well as engineered cages formed by protein assembly \cite{Zhu2021,Edwardson2022}. A significant body of research on densely packing nano-sized particles in cylindrical confinement also exists [e.g.,\cite{landry2003confined,mueller2005numerically,Fu2016}]

Geometric frustration can act on not only one but two types of structural ordering at the molecular level, the physics of which is much less understood.
Competing orders have attracted long-standing interest, e.g., in metallurgy  \cite{Reed1973,Perim2016} and in extraordinary ice structures \cite{Huang2023,Knopf2023}.
This is realized at the molecular level by direct visual observation of softcore particles under an optical microscope \cite{Frenkel2002,Frenkel2006,Manoharan2015,Li2016}.
In quasi-two-dimensional planar systems,
both hexagonal (Hex) and square (Sq) lattices can be stabilized experimentally  \cite{Peng2015,Singh2022,Rey2017,Rossi2015,Zhao2011}.
The Hertzian potential energy, which has
a soft repulsion core below the interaction range to capture the elastic nature
between colloidal particles, has become the choice in  computer simulations \cite{Miller2011,Terao2013,Zu2016,Yao2020}, modeling the stability of Hex, Sq and quasi-crystals when colloidal
particles are
closely packed. It has also been successfully used to model the experimental results of soft microparticles in a flat two-dimensional plane \cite{Grillo2020}
or a three-dimensional space \cite{mohanty2014effective}, unraveling
the physical insights behind the real systems.

To elucidate how geometric frustration affects order-order competitions, in this study, we provide a physical picture of the structural formations on the spherical surface using the Hertzian interaction potential.
Through extensive computer simulations of systems containing
a large number of soft particles in various packing densities, rich
structural configurations are discovered near the Hex-Sq solid-to-solid transition density.
The existence of domain and counter-domain defects is revealed, which were rarely discussed
in the literature.
The competition between Hex- and Sq-lattice stabilities near the solid-solid transition point enables
rectangular defect domains (rich in Sq tiling) in a Hex-tile background, and triangular defect domains (rich in Hex tiling) in a Sq-tile background.
The compound winding numbers as a complete unit of a domain defect
make the entire system to arrange the defect locations differently from those of point or scar defects.

More surprising is the emergence of counter domains
when Hex- and Sq-tiles have comparable area fractions and the defect domains occupy a large surface area.
The edges between the hosting defect domain and the background lattice are
substantially curved, forming a spherical polygon (biangle, triangle, etc.), which has a larger sum of internal angles than their flat-space counterparts, a typical consequence of the Gauss-Bonnet theorem.
Neither Hex- nor Sq-tiles can uniformly
fill the host domain area without creating lattice mismatches. Then,
enclave counter-domain defects appear within the host
domain defects.

The study was performed by using the simulated annealing Langevin dynamics in LAMMPS \cite{LAMMPS} in many independent runs,
to find the low-energy states of $N$ particles confined on the spherical
surface of radius $R$. The Hertzian potential \cite{Landau1970,Johnson1985} is used to model
the interaction between particles having an interaction range $\sigma$.
Two dimensionless
parameters,
$N$ and the reduced packing density
$\rho^* = N\sigma^2/4\pi R^2$, describe the physical state of the problem.
In a flat space, in the thermodynamic limit, one can show that the uniform Hex lattice is stable when
$\rho^*<2.255...$ and Sq lattice when $\rho^*>2.364...$, theoretically.
In a previous computer simulation, the crossing of $\phi_6$ and $\phi_4$, at $\rho^*=2.228$, is thought of
as the first-order Hex-Sq phase transition point \cite{Yao2020}. On a spherical surface, all interesting regimes appear near these densities.

\textbf{State diagram and defects.}
Figure \ref{FIG1}(a) illustrates the state diagram
found from our study, of relatively large systems ($N\geq750$). To find the ground states, we performed independent  20 simulated annealing simulations for each of the 840   uniformly distributed sampling points
 in the state diagram; among these, the states having the lowest energies are chosen and analyzed.
Based %
on the defect patterns %
discussed below,
six distinct regimes are obtained:
(i) the surface is covered by a single, deformed Hex lattice with point disclinations [Fig. \ref{FIG1}(b)] or scars;
(ii) the Hex-rich background encloses Sq-lattice domain defects [Fig. \ref{FIG1}(c)]; 
(iii) the Sq-lattice domain defects in (ii) further enclose Hex-rich domain defects [Figs. \ref{FIG1}(d) and (e)] ;
(iv) non-symmetric patterns where Hex-rich and Sq-rich domains are phase-separated [Figs. \ref{FIG1}(f) and (g)];
(v) Sq-rich background containing scattered  Hex domain defects [Fig. \ref{FIG1}(h)];
and (vi) a single, deformed Sq-lattice background hosting point or scar disclinations [Fig. \ref{FIG1}(i)].

The configurational properties described below
can be repeatedly confirmed from the simulations of all regimes.
The boundary between regimes-(i) and (ii) is not a clear line division; in a narrow corridor
both types of defect patterns can coexist with comparative system energies. %
Along the dashed boundaries, coexistence of the nearby states is also observed, but in
 very narrow density ranges.

The smaller systems ($N\lesssim 750$) of competing lattices display distinct, finite-size behavior,
as the particle numbers significantly affect the morphology of the minimum energy structure.
In Ref.
\cite{Miller2011}, defect structures up to $N=12$ were classified for Hertzian molecules on the
spherical surface, which are all repeated and verified here. Some additional, small-$N$ structures are provided in SI
 \cite{SI}.

In Hex-rich regime-(i), 12 point or scar disclinations, each of winding number $w=+1/6$,
 display an overall icosahedral
symmetry, expected by the Caspar-Klug construction
of hexagonal and pentagonal lattices on a spherical
surface. Conversely, regime-(vi) is dominated by a defect structure
where an ocean of Sq cells hosts eight defect domains rich
in Hex cells or eight scars, of a net $w = +1/4$. These eight
defects form the corners of an anticube. See examples in Figs. \ref{FIG1}(b) and (i). %
The scar properties are examined next.

\textbf{Scars.}
A distinct type of disclination patterns of lattices
on the spherical surface is scars that form
a defect line
dividing two regular lattice regions. Its existence in a real system was
first shown by assembling colloid beads on the surface of a liquid droplet  \cite{Bausch2003}.
Scars are commonly observed in the current system, in the large $N$-regions of Hex and Sq states, starting in about $N=200$.
Point disclinations in low $N$ evolve into scar defects.

Figures \ref{FIG2}(a) and \ref{FIG2}(b) show typical scar patterns in a background of Hex- and Sq-lattices.
A scar is characteristically
composed of a linear arrangement of
complementary point defects, for example,
with winding numbers $(+1/6)$-$(-1/6)$-$
(+1/6)$-$(-1/6)$-$(+1/6)$ in Hex. The net winding number of the entire scar is $+1/6$, same as a typical point defect in a small-$N$ system.
Energetically, each uniform domain of the triangular Hex lattices prefer a regular triangular configuration. As the two uniform domains meet on the spherical surface at an unmatchable angle, however, the
mismatch of the uniformity necessitates a defect line to stitch them together. Note that $w=+1/6$ and $-1/6$ point defects can be
understood as removal and insertion of a regular triangle, respectively. Along the stitching line, triangles are alternatively removed and inserted. This can also be considered as
a 5-7-5-7-5 copolymer chain, referring to the number of bonds along the scar  \cite{Bausch2003}. The thickness of the scar is about one lattice size.

In contrast, a scar in the Sq state observed here has different
features and occurs when two uniform Sq domains meet on the spherical surface at an angle.
It is the manifestation of a disclination having a total $w=+1/4$.
On a convex surface, defects of $w=-1/4$ carry a large energy penalty hence are never preferred. This rules out the existence of a scar of the type $(+1/4)$-$(-1/4)$-$
(+1/4)$-$(-1/4)$-$(+1/4)$, which could be imagined based on a naive comparison with the scar in the Hex state.
Instead, a less energy-cost configuration is to match the weakly distorted squares near the boundary with a
sequence of distorted triangles [see Fig. \ref{FIG2}(b)]. The molecules connecting this triangle sequence are associated with two squares and two triangles, all weakly
distorted, hence forming a backbone of homopolymer of winding number $...(-1/12)$-$(-1/12)$-$
(-1/12)$-$(-1/12)$-$...$ On both sides
of the backbone, three Sq tiles connect to a single triangle, which is a typical
$w=+1/12$ defect point, complementary to $w=-1/12$.
Because of the geometrical mismatch of inserted triangles and connected squares, these
side defects wiggles every couple of monomers to follow a domain boundary smooth at a much larger scale. Hence, the thickness of the scar is approximately 2 lattice size.

In large $N$, the number of excess disclination pairs (i.e., half of the difference between the number of total disclination
on surface and the essential disclinations, 12 for Hex and 24 for Sq.), $N_d$, involved in both types of scars
follows a power-law scaling. Assuming
that the relative scar-length to radius ratio is fixed, one writes $N_d = \beta R/a$,
where $a$ is the bond length and $\beta$ a  numerical coefficient \cite{Bowick2009}.
Then,
\begin{equation}\label{Nddef}
	N_d = \alpha N^{1/2}
\end{equation}
where $\alpha$ is a numerical constant.
This power law is verified by our data in Fig. \ref{FIG2}(c) in both Hex and Sq states up to $N=7200$, at various
densities.
The ratio of the coefficients is $\alpha_{\text {Hex}}/
\alpha_{\text {Sq}}=(1.34\pm 0.03)/( 3.5 \pm 0.2) = 0.38 \pm 0.02$.
Taking into account the area ratio between triangular and
Sq cells, one also expects
$\alpha_{\text {Hex}}/\alpha_{\text {Sq}} =
(3^{1/4}/2)
(\beta_{\text {Hex}}/\beta_{\text {Sq}})$
where $\beta_{\text {Hex}}/\beta_{\text {Sq}}=0.544$
was deduced from a generic lattice model \cite{Bowick2009}.
The %
agreement indicates that the ratio $\alpha_{\text {Hex}}/
\alpha_{\text {Sq}}$ is universal, independent of the microscopic details of the molecular potential energy.

\textbf{Shape frustrations, domains, and  counter domains.} In regimes  (ii)-(v),
Hex and Sq domains both occupy comparably large surface areas, preferred by phase separation. The
curvature effects on the
boundaries then become significant.
Some display maze-like mixing, interwinding Hex- and Sq-rich bands without well-defined domains [e.g., Fig. \ref{FIG1}(f)],
 and others counter domains,
where one type of lattice resides inside a domain of another
type of lattice that by itself is a defect of a larger
background lattice of the first type.
 The existence of the maze-like, connected structures was recently observed in experiments of suspending molecules \cite{Das2022} and growth of two dimensional rigid colloidal lattices \cite{Meng2014} on the spherical surface of droplets.

The separation between domains rich in Hex- and Sq-lattices is commonly observed in regimes-(ii), (iii), and  (v) of the state diagram. Typically, patches of localized Sq-lattice domains appear in regime-(ii), in the background of Hex tiles, where within
the relatively flat local area, the domain boundaries mostly follow straight lines [see Fig. \ref{FIG1}(c)].
The overall $4\times(+1/12)=+4/12$ is then the winding number of
a rectangular domain of Sq lattice in the Hex background.
An idealized structure is shown in Fig. S2(c) for $N=122$ and $\rho^*=2.27$, where six such Sq
domains are located at the vertices of an octahedron, which maximize the distances from each other on spherical surface.

In regime-(v),
now of higher $\rho^*$, where Sq order becomes dominated with larger area fractions, localized equilateral-triangle domains of Hex lattice
are formed with rather
straight edges.
An example can be seen in Fig. \ref{FIG1}(h), for $N=1000$ and $\rho^*=2.35$.
The overall triangular domain hence carries the winding number $3\times (+1/12) = + 3/12$ collectively.
Eight triangular domains exist, forming the
corners of a cube. In other examples, the eight triangular domains can also display twisted-cube formation on the
spherical surface [Fig. S2(k)].

To demonstrate that the appearance of counter domains is an imperative
consequence of the curved nature of the spherical surface,
the structure in Fig. \ref{FIG1}(d) is used as the first example.
From geometry characteristics, the Gauss-Bonnet theorem dictates that the spherical triangle covering one quadrant
has a corner angle  of $\pi/2$, hence the square pattern is the natural choice of tiling the near-edge interior [see the right plot of the Fig. \ref{FIG1}(d)].
Though the tiling texture can endorse weak deformation moving to the central region
of the domain, the central domain region
becomes a regular triangle, where the corner angle approaches $\pi/3$, hence
the system
must  switch to a triangle based pattern, %
the Hex tiling.
From the physics of energy minimum,
accompanying the Sq lattice deformation inside the spherical triangle,
vertex energies begin to build up in the near-center area. The transformation to the Hex lattice
releases this energy by moving the energy penalty to
singular sites --- in this case
the three defect points at the corners of the enclave domain.

The next example,  in Fig.~\ref{FIG1}(e), a typical metastable state observed at $\rho^*=2.30$,   demonstrates that the shape frustration can yield
an
even more complicated domain morphology, where
not only an enclave domain exists, but another domain is further embedded inside. %
The spherical lune shown in Fig. \ref{FIG1}(e) extends a large area, making $\pi/2$ biangles.
As the interior of the spherical lune, having two right angles at the corner,
is filled by Sq tiles from the edges,
an intermediate spherical
biangle is reached, having $\pi/3$ biangles, which is suited for the Hex tiles.
Hence an enclave, Hex-tiled counter-domain
defect must exist.
Accommodating the reduction of the lunar edge lengths,
two other Hex-tiled domains in the middle are produced simultaneously, in consistency with the Burgers vector
formula  \cite{Chaikin1995}.
As the
central area of the spherical lune is reached,
the spherical lune shrinks to a horizontal lattice line,
which necessitates the existence of square lattices in the center
as another level of
enclave defect domain.
This scenario of defect
domain, counter domain, and counter-counter domain is hence intrinsically required by the
shape frustration on a curved surface.

A winding number analysis indicates that a complex
built from
the hosting and counter domains
in Fig. \ref{FIG1}(d) has $w=+1/2$ and in \ref{FIG1}(e) $w=+2/3$.
The four-domain complexes in Fig. \ref{FIG1}(d) (three are blocked behind)
form  a tetrahedron symmetry on the spherical surface.
This particular pattern can be found in regime-(iii) in Fig. \ref{FIG1}(a).
The three-domain complexes in Fig. \ref{FIG1}(e) are arranged around a big circle, with the same shape
orientations, showing a trigonal hosohedron symmetry on the surface.

Overall,
the final, global configuration that these domain structures settle down, is the manifestation of solid-solid coexistence, shape frustration, and energy minimization.
It can have asymmetric domains and counter domains,
some of which are illustrated in Figs. \ref{FIG1}(f) and (g).

\textbf{Conclusions.} We investigated the impact of a curved surface on the
formation of the crystalline structures by confining $N$ soft
colloid particles
on the spherical surface of radius $R$.
In the vicinity
of the transition density, spherical areas made of hexagon and square lattices
have the tendency to co-exist because of the close matching in lattice energies.
The curved nature of these spherical shapes introduces another
level of frustration and competitions in tiling
regular hexagon and square lattices. As the result, large defect domains
necessitate the existence of  counter domains in the interior and can even establish
another level of counter domains within a counter domain.
The disclination patterns discussed here combine
the physics of topological defects, shape
frustration, lattice energetics, and lattice tiling. Any single
factor can be hardly used to explain the diversity of the observed surface features.
The values of the 
packing densities in the current study depends on the model used;
generally, the domain and counter-domain defects in real systems ought to appear between
their own Hex- and Sq-stable densities.

The computer simulation of interacting particles offers a platform
closely matching  those in real systems. Our predictions may be realized experimentally by extending the competing Hex-Sq lattices from planar systems \cite{Peng2015,Singh2022,Rey2017,Rossi2015,Zhao2011} to spherical systems, e.g., particles confined in double emulsions \cite{Zarzar2015,Lee2008} or rounded cubes on the surface of an emulsion \cite{Pang2013}. The topological-defect problems find their roots in
other topic areas in science: for example, metallurgy  \cite{Reed1973}, molecular
network \cite{Blunt2008}, living matter \cite{Ardavseva2022}, and superconductivity \cite{Liu2019}.
The disclination patterns discussed here provide a unique opportunity to investigate the
combined physics of topological defects, shape
frustration, lattice energetics, and lattice tiling.

\begin{acknowledgement}
This work was supported by the National Natural Science Foundation of China (12275137), Fundamental Research Funds for the Central Universities, Nankai University (63221053, 63231190), and Natural Sciences and Engineering Council of Canada. We thank Fangfu Ye, Weichao Shi, and Baohui Li for discussion, and ComputeCanada for providing computational
resources.

\textit{Author Contributions.} Y.L. designed the project. H.X, W.L and Z.L. performed the simulations. H.X., W.L, Z.L., J.Z.Y.C, and Y.L. analyzed the data and wrote the manuscript.

\textit{Competing interests.} The authors declare no competing interests.

\end{acknowledgement}

\begin{suppinfo}

The following files are available free of charge.
\begin{itemize}
    \item Simulation methods (SI A), Bond order parameter (SI B), Defects and winding numbers (SI C and Figure S1), Defect structures (SI D and Figure S2), Winding numbers for Fig. 1(d) and Fig. 1(e) in the text. (SI E)
\end{itemize}

\end{suppinfo}

\providecommand{\latin}[1]{#1}
\makeatletter
\providecommand{\doi}
{\begingroup\let\do\@makeother\dospecials
	\catcode`\{=1 \catcode`\}=2 \doi@aux}
\providecommand{\doi@aux}[1]{\endgroup\texttt{#1}}
\makeatother
\providecommand*\mcitethebibliography{\thebibliography}
\csname @ifundefined\endcsname{endmcitethebibliography}
{\let\endmcitethebibliography\endthebibliography}{}

\end{document}